\documentclass[9pt,conference]{IEEEtran}
\usepackage{dcase2025}
\usepackage{bm} 
\usepackage{amsmath,graphicx,xurl,times,booktabs, tabularx,comment}

\usepackage{hhline}
\usepackage{hyperref}
\usepackage{xcolor}
\usepackage{cite}
\usepackage{multirow,url,hyperref}
\usepackage{siunitx}

\usepackage{pifont}
\usepackage{multirow}
\usepackage{comment}

\usepackage{hyperref}
\usepackage{xcolor}

\usepackage{etoolbox}
\makeatletter
\pretocmd{\thebibliography}{\footnotesize}{}{}
\makeatother

\title{Low-Complexity Acoustic Scene Classification with Device Information \\ in the DCASE 2025 Challenge}

\name{Florian Schmid$^{1}$,
       Paul Primus$^{1}$,
       Toni Heittola$^{2}$,
       Annamaria Mesaros$^{2}$,
       Irene Martín-Morató$^{2}$,
       Gerhard Widmer$^{1,3}$
       }
 \address{$^1$Institute of Computational Perception, Johannes Kepler University Linz, Austria \\
 $^2$Computing Sciences Tampere University, Finland, $^3$LIT Artificial Intelligence Lab, Linz, Austria\\
 \{florian.schmid, paul.primus, gerhard.widmer\}@jku.at\\
 \{toni.heittola, annamaria.mesaros, irene.martinmorato\}@tuni.fi
  }

\begin{document}

\maketitle

\begin{sloppy}

\begin{abstract}

This paper presents the \textit{Low-Complexity Acoustic Scene Classification with Device Information} Task of the DCASE 2025 Challenge, along with its baseline system. Continuing the focus on low-complexity models, data efficiency, and device mismatch from previous editions (2022–2024), this year's task introduces a key change: recording device information is now provided at inference time. This enables the development of device-specific models that leverage device characteristics—reflecting real-world deployment scenarios in which a model is designed with awareness of the underlying hardware. The training set matches the 25\% subset used in the corresponding DCASE 2024 challenge, with no restrictions on external data use, highlighting transfer learning as a central topic. The baseline achieves 50.72\% accuracy with a device-agnostic model, improving to 51.89\% when incorporating device-specific fine-tuning. The task attracted 31 submissions from 12 teams, with 11 teams outperforming the baseline. The top-performing submission achieved an accuracy gain of more than 8 percentage points over the baseline on the evaluation set.

\end{abstract}


\section{Introduction}
\label{sec:intro}

Acoustic Scene Classification (ASC) aims to identify the type of environment in which an audio recording was made, based on a short excerpt~\cite{benetos2018approaches}. Environments are defined as a set of real-world locations, such as \textit{Metro station}, \textit{Urban park}, or \textit{Public square}. The ASC task has a long-standing presence in the DCASE Challenge, evolving through various refinements over the years. Recent editions have emphasized challenges relevant to real-world deployment, including low-complexity constraints~\cite{heittola20asc_in_dcase20, Martin21asc_in_dcase21, Martin22asc_in_dcase22,schmid24asc_in_dcase24}, recording device mismatch~\cite{Mesaros18multi_dev_dataset, heittola20asc_in_dcase20,schmid24asc_in_dcase24}, and data efficiency~\cite{schmid24asc_in_dcase24}. For example, the 2024 edition required systems to be lightweight enough to operate on embedded devices, to achieve high performance with limited training data, and to generalize across a variety of potentially unknown recording devices. The 2025 edition\footnote{Task Description Page: \href{https://dcase.community/challenge2025/task-low-complexity-acoustic-scene-classification-with-device-information}{https://dcase.community/challenge2025/task-low-complexity-acoustic-scene-classification-with-device-information}}
 introduces several modifications compared to the 2024 edition. The most significant change in the 2025 edition is the availability of the recording device ID at inference time. This enables participants to tailor their models to device-specific characteristics, for instance, by fine-tuning the model for the known hardware. This design reflects realistic deployment scenarios where the target device is known in advance and recordings from it may be available to improve prediction accuracy. 

\begin{figure}[t!]
\centering
{\includegraphics[width=\linewidth]{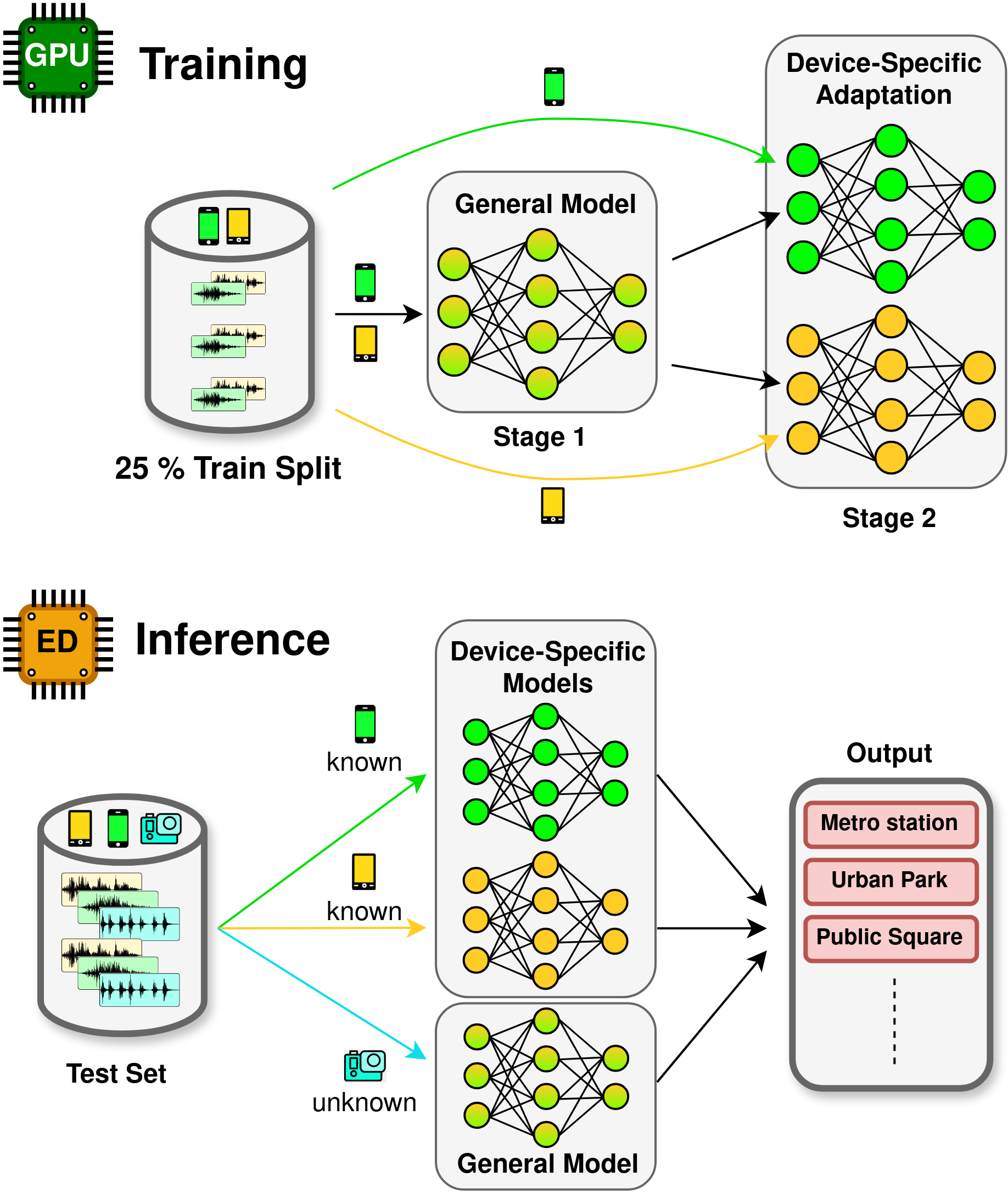}}
\caption{Overview of \textit{Low-Complexity Acoustic Scene Classification with Device Information}. At inference time, models must operate under low-complexity constraints and handle both known (seen during training) and unknown (unseen during training) recording devices, with the device ID provided. The baseline follows a two-stage training process: first, learning a general model, then adapting it to device-specific characteristics to enhance performance on known devices.}
\label{fig:overview}
\end{figure}

Figure~\ref{fig:overview} illustrates the task setup and baseline training procedure. Training is performed in two stages: a \textit{general model} is first trained on the full available dataset (25\% subset from the 2024 edition), followed by adaptation into \textit{device-specific models} using recordings from known devices. At inference, \textit{device-specific models} are used for known devices, while the \textit{general model} handles unknown ones. All models must comply with the low-complexity constraints, ensuring suitability for embedded devices (ED).

The limited size of the training set reflects real-world scenarios with scarce labeled data, highlighting transfer learning as a key strategy. In contrast to 2024, the 2025 task lifts restrictions on external resources, allowing participants to incorporate additional acoustic scene datasets to improve performance.

The remainder of the paper is organized as follows: Section~\ref{sec:previous} briefly reviews prior approaches to device generalization, low-complexity constraints, and transfer learning in earlier challenge editions. Section~\ref{sec:setup} details the task setup, and Section~\ref{sec:bl_system} presents the baseline system. Results are discussed in Section~\ref{sec:results}, and conclusions are drawn in Section~\ref{sec:conclusion}.

\section{Previous Editions}
\label{sec:previous}

In past editions, several strategies were proposed to improve generalization across different—and potentially unknown—recording devices. The most common in 2023 and 2024 were augmentation-based methods, such as Freq-MixStyle~\cite{Kim2022FMS, Schmid2022CPJKU22} and device impulse response augmentation~\cite{Morocutti23DIR}. Others aimed to suppress device information via domain adaptation~\cite{Truchan25da, Koutini2020CPJKU20} or normalization~\cite{Kim2021qti}, while a third line of work adjusted the sampling distribution to balance devices~\cite{Lee2022}.

Over the years, various complexity constraints have been introduced, with the two most recent editions limiting model size to 128 kB and computational cost to 30 million multiply-accumulate operations (30 MMACs), targeting Cortex-M4-class devices. In response, techniques such as Knowledge Distillation~\cite{Schmid2022CPJKU22}, Pruning~\cite{Koutini2021CPJKU21, Han2024rank1}, and Sparsification~\cite{Yang2021lottery} were explored, alongside the design of efficient CNN architectures~\cite{Tan2023a, Cai2023a, Schmid2023CPJKU23, Han2024rank1, Shao2024rank2}.

To tackle data scarcity, the 2024 edition saw widespread use of transfer learning from the large-scale general-purpose audio dataset \textit{AudioSet}~\cite{Gemmeke17audioset}. Participants leveraged it in three main ways: (1) fine-tuning a large pre-trained model on ASC and distilling it into a low-complexity student\cite{Han2024rank1, Shao2024rank2, Cai2024rank5}; (2) pre-training a low-complexity model directly on AudioSet~\cite{David2024rank2}; or (3) extracting task-relevant clips from AudioSet for training~\cite{Werning2024rank8}.

\section{Task Setup}
\label{sec:setup}

As discussed in the previous section, device mismatch, low-complexity constraints, and transfer learning have been extensively studied in the context of the ASC task. However, this year's setup introduces key variations to the handling of device mismatch and transfer learning. Regarding device mismatch, the recording device ID is now provided at inference time. Some device IDs may already have appeared in the training data, others may be novel. This will allow participants to develop specialized models for devices known from the training set. For transfer learning, external datasets are no longer limited to general-purpose collections like AudioSet~\cite{Gemmeke17audioset}. However, related acoustic scene datasets are now permitted. Given these changes, the challenge aims to address the following set of research questions:

\begin{itemize}
    \item Can device type information be exploited to improve performance compared to previous editions, where it was not available at inference time?
    \item Which machine learning techniques are most effective for creating specialized models for different recording devices?
    \item Can additional acoustic scene datasets—possibly featuring different scenes, locations, or devices—help improve performance on the TAU dataset~\cite{Mesaros18multi_dev_dataset, heittola20asc_in_dcase20}?
\end{itemize}

\subsection{Dataset}

The task again builds on top of the \textit{TAU Urban Acoustic Scenes 2022 Mobile} dataset~\cite{Mesaros18multi_dev_dataset, heittola20asc_in_dcase20}, which was also used in the 2022, 2023, and 2024 editions of the challenge~\cite{Martin22asc_in_dcase22,schmid24asc_in_dcase24}.
The dataset provides one-second audio snippets sampled at 44.1 kHz in single-channel, 24-bit format and consists of recordings from ten distinct acoustic scenes.

Audio was captured in multiple European cities using four devices in parallel: a high-quality binaural recorder (primary device \textit{A}) and three consumer devices (\textit{B}, \textit{C}, \textit{D}). Additionally, ten simulated devices (\textit{S1}–\textit{S10}) were created by applying device-specific impulse responses to recordings from device A. For further details on the dataset creation and device distribution, we refer to~\cite{heittola20asc_in_dcase20}. This dataset description is based on~\cite{schmid24asc_in_dcase24}. The dataset is divided into a \textit{development set} and an \textit{evaluation set}, following a predefined split.

\textbf{Development Set:}  
The development set contains 64 hours of audio recorded with three real devices (A, B, C) and six simulated devices (S1--S6). It is further divided into:
\begin{itemize}
    \item \textit{Development-train}: This corresponds to the 25\% subset used in last year's data-efficient evaluation setup~\cite{schmid24asc_in_dcase24}. It includes recordings from six devices: A, B, C, and S1--S3.
    \item \textit{Development-test}: In addition to the devices in development-train, this split includes the remaining simulated devices S4--S6, which are unseen during training and serve to evaluate generalization to unknown devices.
\end{itemize}

Only the development-train split (25\% subset) and announced external resources may be used for training. The development-test split must be used only for evaluation. City and device information are provided for all recordings in the development set.

\begin{table*}[t]
\centering
\caption{Device-wise and overall accuracies of the baseline system on the development-test split. }
\label{tab:baseline_results}
\begin{tabularx}{\textwidth}{l*{9}{c}c}
\toprule
\textbf{Model} & \textbf{A} & \textbf{B} & \textbf{C} & \textbf{S1} & \textbf{S2} & \textbf{S3} & \textbf{S4} & \textbf{S5} & \textbf{S6} & \textbf{Macro Avg. Accuracy} \\
\midrule
General Model           & 62.80 & 52.87 & 54.23 & 48.52 & 47.29 & 52.86 & 48.14 & 47.23 & 42.60 & $50.72 \pm 0.47$ \\
Device-specific Models  & 63.98 & 55.85 & 59.09 & 48.68 & 48.74 & 52.72 & 48.14 & 47.23 & 42.60 & \textbf{51.89 $\pm$ 0.05} \\
\bottomrule
\end{tabularx}
\end{table*}

\textbf{Evaluation Set:}
The evaluation set includes five unknown devices (D and S7--S10), as well as two cities that are not present in the development set, in addition to recordings from known cities and devices. It is used for final system evaluation and is published without scene labels. Device IDs are provided at inference time, while city information is withheld. Known devices (A, B, C, S1--S3) are labeled explicitly, whereas unknown devices (D, S7--S10) are marked as \textit{unknown}. The ratio of known to unknown devices is kept consistent between the development-test and evaluation sets.

\subsection{Device-Specific Modeling: Problem Setting}

In this section, we briefly formalize the problem setting arising from the availability of device information.
We assume the training data is drawn from $K$ distinct domains (i.e., devices) ${D_1, D_2, \dots, D_K}$, each associated with its own data distribution $p_{D_k}(X)$. The amount of training data per domain varies and is often limited. The domain ID is provided with each training example.

At test time, the system is evaluated on samples originating from a mix of \textit{known} domains (seen during training) and \textit{unknown} domains (unseen during training). For each test sample, the corresponding source domain (i.e., device ID) is provided. This additional information allows for models that specialize in known domains by leveraging domain-specific characteristics, while still requiring a general model to handle unknown domains.  

A straightforward strategy to address this setting is to first train a general model across all domains and then adapt it to individual domains using the corresponding in-domain training data. This two-step approach is also implemented in the baseline system, as described in Section~\ref{sec:bl_system}. Key innovations may lie in the strategy for specializing the general model to the known domains, which may contain only a small number of labeled data points.

\subsection{Evaluation and Submission}

Submissions are ranked based on class-wise macro-averaged accuracy computed on the evaluation set. As a secondary, operating point-independent metric, multi-class cross-entropy is reported. Each team may submit up to four sets of predictions from different systems. This year, participants must also submit inference code to promote open research and allow additional complexity evaluations by the organizers.

\subsection{System Complexity Requirements}
\label{subsec:complexity}

The system complexity constraints follow the 2024 edition~\cite{schmid24asc_in_dcase24} and apply to each individual model, including both the general model and any device-specific variants. Both model size and computational cost are restricted. Specifically, model parameters must fit within 128~kB of memory, with no fixed numerical precision requirement. Participants are free to trade off the number of parameters against numerical precision; for instance, the limit corresponds to 128K parameters with 8-bit quantization or 32K parameters with 32-bit precision. Computational complexity is capped at 30~MMACs for processing a one-second audio segment. These constraints are designed to reflect the capabilities of resource-constrained devices such as the Cortex-M4 series (e.g., STM32L496@80\,MHz or Arduino Nano 33@64\,MHz).

\section{Baseline System}
\label{sec:bl_system}

Following the 2024 edition~\cite{schmid24asc_in_dcase24}, the baseline system builds on a simplified variant of the top-performing submission from the 2023 edition~\cite{Schmid2023cpm}. It employs a receptive-field-regularized, factorized CNN architecture, referred to as \textit{CP-Mobile}. Audio recordings are first resampled to 32~kHz, then converted into mel spectrograms using a 4096-point FFT with a window size of 96~ms and a hop size of approximately 16~ms, followed by a mel scaling with 256 mel filterbanks.

As illustrated in Figure~\ref{fig:overview}, the system is trained in two stages. In the first stage, a \textit{general model} is trained on data from all devices for 150 epochs using the AdamW optimizer and a batch size of 256. To address device mismatch, Freq-MixStyle~\cite{Kim2022FMS,Schmid2022CPJKU22} is applied during training. In the second stage, for each device in the training set, a \textit{device-specific model} is created by end-to-end fine-tuning the \textit{general model} on data from that specific device for 50 epochs. During inference, device-specific models are applied to known devices, while the general model handles unknown ones.

The baseline system requires 29.4~MMACs to process a one-second audio clip. The model uses 61,148 parameters in 16-bit (fp16) precision, resulting in a total memory footprint of 122.3~kB for the parameters.

Table~\ref{tab:baseline_results} presents the device-wise and overall accuracies of the baseline system on the development-test split. After Stage~1, the \textit{general model} achieves an overall accuracy of 50.72\%. Following Stage~2, where device-specific models are trained, the overall accuracy improves to 51.89\%. Device-specific fine-tuning increases the accuracy for all known devices except for S3, with performance gains varying notably across devices. The accuracy on unknown devices remains unchanged between the two rows of the table, as the \textit{general model} is used for inference on unknown devices. The source code and a detailed description of the baseline system are available online\footnote{Source Code: \href{https://github.com/CPJKU/dcase2025_task1_baseline}{https://github.com/CPJKU/dcase2025\_task1\_baseline}}.

\begin{table*}[t]
\centering
\caption{Best-performing system per team (only including systems that outperform the baseline) and the official DCASE2025 baseline. \textbf{Score} indicates the accuracy on the evaluation set, \textbf{Size} refers to the memory required to store model parameters, and \textbf{MAC} denotes the number of multiply-accumulate operations. \textbf{External} indicates whether external data was used, and \textbf{Device Adaptation} describes the method used to adapt the model to specific devices based on provided device IDs. \textbf{KD}, \textbf{IR}, and \textbf{FT} stand for Knowledge Distillation, Impulse Response augmentation, and Fine-Tuning, respectively.}
\label{tab:results}
\begin{tabular}{@{}lcccccccc@{}}
\toprule
\textbf{Team} & \textbf{Score} & \textbf{Size} & \textbf{MAC} & \textbf{Architecture} & \textbf{Complexity} & \textbf{External} & \textbf{Device Adaptation} \\
\midrule
Karasin\_JKU         & 61.5 & 122kB & 29M & CP-Mobile & fp16,KD & IR,CochlScene,BEATs & Full device-spec. FT \\
Tan\_SNTLNTU         & 59.9 & 116kB & 10M & CNN-GRU & fp16, prune & IR & Full FT\\
Luo\_CQUPT           & 59.6 & 123kB & 28M & DynaCP & fp16, KD & EfficientAT & Full FT \\
Zhang\_AITHU-SJTU    & 59.3 & 126kB & 29M & SSCP-Mobile & fp16,KD,prune & PaSST & -- \\
Chang\_HYU           & 59.0 & 125kB & 29M & Rep-CTFA & fp16,KD & IR,PaSST & Head-only FT\\
Li\_NTU              & 58.9 & 122kB  & 17M & CP-Mobile & KD,prune & IR,PaSST & -- \\
Ramezanee\_SUT              & 57.9 & 125kB & 28M &  DSFlexiNet & KD  & IR & Full FT \\
Jeong\_SEOULTECH     & 57.9 & 122kB  & 26M & CP-Mobile & fp16,KD & PaSST & Full FT \\
Chen\_GXU            & 56.6 & 122kB & 29M & CP-Mobile & fp16,KD & PaSST & -- \\
Krishna\_SRIB   & 56.1 & 122kB  & 27M & CP-Mobile & fp16 & -- & Full FT \\
Zhou\_XJTLU   & 55.5 & 126kB  & 29M & TF-SepNet & int8,KD & IR,AudioSet,BEATs & Full FT \\
\midrule
\textbf{DCASE25 baseline} & \textbf{53.2} & 122kB & 29M & CP-Mobile & fp16 & -- & Full FT \\
\bottomrule
\end{tabular}
\end{table*}

\section{Challenge Results}
\label{sec:results}

The task received 31 submissions from 12 teams, with 11 out of 12 teams outperforming the baseline system. For both the baseline and most submitted systems, performance on the development-test split aligned well with that on the evaluation set. Table~\ref{tab:results} presents the best-performing system from each team that outperformed the baseline and summarizes their architectural choices, strategies for handling complexity, use of external data, and device adaptation methods. The following subsections discuss each of these aspects in detail. Additional results and detailed system descriptions are available on the official challenge website\footnote{\href{https://dcase.community/challenge2025/task-low-complexity-acoustic-scene-classification-with-device-information-results}{Results: https://dcase.community/challenge2025/task-low-complexity-acoustic-scene-classification-with-device-information-results}}.

\subsection{Architectures}

Due to the low-complexity constraints, efficient neural network design remained a central focus. In line with last year’s trends~\cite{schmid24asc_in_dcase24}, most teams adopted factorized convolutional architectures. Five of the twelve teams—including the top-ranked submission—built their systems on the CP-Mobile architecture~\cite{Schmid2023cpm}. However, several top-performing teams proposed novel architectural variants.

Team \textit{Tan\_SNTLNTU}~\cite{Tan2025rank2} introduced \textit{CNN-GRU}, which combines pointwise and 1D depthwise convolutions over the frequency and time dimensions, integrates Squeeze-and-Excitation layers~\cite{Hu18SE}, and applies a GRU along the frequency axis. Team \textit{Luo\_CQUPT}~\cite{Luo2025rank3} presented \textit{DynaCP}, a CP-Mobile modification that processes pooling and strided convolutions in parallel and dynamically combines their outputs. Teams \textit{Chang\_HYU}~\cite{Chang2025rank5} and \textit{Ramezanee\_SUT}~\cite{Ramezanee2025rank7} built upon reparameterizable convolution blocks~\cite{Han25rank1_icassp}, which use multiple branches during training that can be merged into a single, efficient equivalent at inference time. Additionally, \textit{Chang\_HYU}~\cite{Chang2025rank5} employed Channel-Time-Frequency Attention (CTFA)~\cite{zeng2025ctfa}, a lightweight attention mechanism that allows the model to focus on informative input regions, while \textit{Ramezanee\_SUT}~\cite{Ramezanee2025rank7} proposed learnable pooling layers. As input to the models, all teams used log-mel energies, with the exception of two teams that used the spectrogram instead.

%

\subsection{System Complexity}

As in previous editions~\cite{schmid24asc_in_dcase24,Martin22asc_in_dcase22}, Knowledge Distillation (KD)~\cite{Hinton15kd} remained the most widely used model compression technique, employed by 10 out of 12 teams. Compared to previous editions, several interesting variations to the KD process have been explored, such as a feature-level distillation loss~\cite{Li2025rank6}, device-aware feature alignment loss to train a device-expert teacher~\cite{Jeong2025rank8}, and self-distillation~\cite{Chen2025rank9}.

Compared to 2024~\cite{schmid24asc_in_dcase24}, where pruning was used only by the top-ranked team~\cite{Han2024rank1}, this year pruning gained traction, with 3 of the top 6 teams adopting it. Notably, the second-ranked team, \textit{Tan\_SNTLNTU}~\cite{Tan2025rank2}, applied pruning exclusively, without using KD. All top-5 teams used 16-bit precision, while none opted for 8-bit quantization—likely due to the ease of reducing to 16-bit with minimal or no accuracy loss, whereas maintaining performance with 8-bit quantization remains more challenging.

\subsection{External Data Usage}

External data was primarily used in two ways. First, most teams employed teacher models for KD that were pre-trained on AudioSet~\cite{Gemmeke17audioset}. PaSST~\cite{koutini22passt} remained a popular choice, though two teams—including the top-ranked one—used BEATs~\cite{chen23beats}, while the third-ranked team, \textit{Luo\_CQUPT}~\cite{Luo2025rank3}, used AudioSet-pretrained MobileNets~\cite{Schmid23EfficientAT} and Dynamic MobileNets~\cite{Schmid24DyMN}.

Second, several teams applied Device Impulse Response (DIR) augmentation~\cite{Morocutti23DIR} using impulse responses from MicIRP\footnote{https://micirp.blogspot.com/}, increasing the diversity of recording conditions in the training data.

Among all participating teams, only the top-ranked team, \textit{Karasin\_JKU}~\cite{Karasin2025rank1}, took advantage of the new rule allowing external ASC datasets by leveraging CochlScene~\cite{CochlScene}. CochlScene contains 76,115 ten-second audio clips recorded across 13 distinct acoustic scenes. The dataset was collected via crowdsourcing, primarily from contributors in Korea. Several scene classes overlap partially with those in the \textit{TAU Urban Acoustic Scenes 2022 Mobile} dataset~\cite{Mesaros18multi_dev_dataset,heittola20asc_in_dcase20} (e.g., \textit{Bus} and \textit{Park}), though others are unique to CochlScene (e.g., \textit{Restroom} and \textit{Elevator}) or TAU (e.g., \textit{Airport} and \textit{Travelling by Tram}).

Team \textit{Karasin\_JKU}~\cite{Karasin2025rank1} explored pre-training both the teacher and student models on CochlScene. Notably, this strategy led to substantial performance improvements for convolutional architectures such as CP-Mobile~\cite{Schmid2023cpm} and CP-ResNet~\cite{koutini21cpr}, with gains of +3.36 and +6.05 percentage points on the TAU development-test split, respectively. In contrast, transformer-based models like PaSST~\cite{koutini22passt} and BEATs~\cite{chen23beats} saw only marginal or no improvements.

\subsection{Device Adaptation}

To exploit the given device information, most teams opted for the baseline strategy of fine-tuning the general model on device-specific data to obtain specialized models. More advanced methods were explored by only a few participants.

Team \textit{Han\_CSU}~\cite{Han2025rank12} addressed device variability by incorporating device embeddings into the model's internal representations, effectively conditioning the network on the identity of the recording device.

Team \textit{Chang\_HYU}~\cite{Chang2025rank5} adopted a modular approach by training lightweight, device-specific classification heads while keeping the shared backbone frozen. This design preserves a common, general-purpose acoustic feature extractor across all devices, while allowing for device-tailored classification at the output stage. Importantly, this method keeps the overall system compact, as the additional device-specific components introduce only minimal overhead.

The top-ranked team, \textit{Karasin\_JKU}~\cite{Karasin2025rank1}, further exploited device information by customizing training configurations—such as Knowledge Distillation hyperparameters—for each device-specific fine-tuning run. In particular, they observed that the optimal loss weighting factor in Knowledge Distillation, which balances the supervised loss and the distillation loss, varies across devices and benefits from device-specific tuning.


\section{Conclusion}
\label{sec:conclusion}

This paper introduced the setup and baseline system for Task~1 of the DCASE~2025 Challenge, which continues to address three core challenges of acoustic scene classification: low-complexity constraints, device mismatch, and limited training data. A key novelty this year is the availability of device information at inference time, enabling device-specific adaptation and yielding consistent improvements in the baseline system.

While the three research questions outlined in Section~\ref{sec:setup} remain only partially explored, the top-ranked submission provided valuable initial answers. They showed that fine-tuning routines tailored to specific devices improve performance, and that leveraging external acoustic scene classification datasets such as CochlScene can substantially boost accuracy on the TAU dataset. These strategies delivered an accuracy gain of more than 1.5 percentage points over all other submissions, highlighting promising directions for future work.

Beyond transfer learning and device-aware modeling, participants also advanced research on efficient architectures, Knowledge Distillation, and pruning. Several teams experimented with different teacher models for Knowledge Distillation, while others introduced architectural components for low-complexity models such as lightweight attention mechanisms, reparameterizable convolutions, and learnable pooling layers.

Overall, the 2025 edition of Task~1 advanced established research on low-complexity modeling while providing initial insights into device-aware adaptation and the use of external acoustic scene datasets, laying the groundwork for further exploration in these directions.




\section{ACKNOWLEDGMENT}
\label{sec:ack}

The LIT AI Lab is supported by the Federal State of Upper Austria.  Gerhard Widmer's work is supported by the European Research Council (ERC) under the European Union's Horizon 2020 research and innovation programme, grant agreement No 101019375 (Whither Music?). Annamaria Mesaros's work is supported by Academy of Finland grant 332063 ``Teaching machines to listen".


\bibliographystyle{IEEEtran}
\bibliography{refs}

\begin{thebibliography}{10}
\providecommand{\url}[1]{#1}
\def\UrlFont{\rmfamily}
\providecommand{\newblock}{\relax}
\providecommand{\bibinfo}[2]{#2}
\providecommand\BIBentrySTDinterwordspacing{\spaceskip=0pt\relax}
\providecommand\BIBentryALTinterwordstretchfactor{4}
\providecommand\BIBentryALTinterwordspacing{\spaceskip=\fontdimen2\font plus
\BIBentryALTinterwordstretchfactor\fontdimen3\font minus \fontdimen4\font\relax}
\providecommand\BIBforeignlanguage[2]{{%
\expandafter\ifx\csname l@#1\endcsname\relax
\typeout{** WARNING: IEEEtran.bst: No hyphenation pattern has been}%
\typeout{** loaded for the language `#1'. Using the pattern for}%
\typeout{** the default language instead.}%
\else
\language=\csname l@#1\endcsname
\fi
#2}}

\bibitem{benetos2018approaches}
E.~Benetos, D.~Stowell, and M.~D. Plumbley, ``Approaches to complex sound scene analysis,'' in \emph{Cham: Springer International Publishing}, 2018.

\bibitem{heittola20asc_in_dcase20}
T.~Heittola, A.~Mesaros, and T.~Virtanen, ``Acoustic scene classification in {DCASE} 2020 challenge: Generalization across devices and low complexity solutions,'' in \emph{DCASE Workshop}, 2020.

\bibitem{Martin21asc_in_dcase21}
I.~Mart{\'{\i}}n{-}Morat{\'{o}}, T.~Heittola, A.~Mesaros, and T.~Virtanen, ``Low-complexity acoustic scene classification for multi-device audio: Analysis of {DCASE} 2021 challenge systems,'' in \emph{DCASE Workshop}, 2021.

\bibitem{Martin22asc_in_dcase22}
I.~Mart{\'{\i}}n{-}Morat{\'{o}}, F.~Paissan, A.~Ancilotto, T.~Heittola, A.~Mesaros, E.~Farella, A.~Brutti, and T.~Virtanen, ``Low-complexity acoustic scene classification in {DCASE} 2022 challenge,'' in \emph{DCASE Workshop}, 2022.

\bibitem{schmid24asc_in_dcase24}
F.~Schmid, P.~Primus, T.~Heittola, A.~Mesaros, I.~Mart{\'{\i}}n{-}Morat{\'{o}}, K.~Koutini, and G.~Widmer, ``Data-efficient low-complexity acoustic scene classification in the {DCASE} 2024 challenge,'' in \emph{DCASE Workshop}, 2024.

\bibitem{Mesaros18multi_dev_dataset}
A.~Mesaros, T.~Heittola, and T.~Virtanen, ``A multi-device dataset for urban acoustic scene classification,'' in \emph{DCASE Workshop}, 2018.

\bibitem{Kim2022FMS}
B.~Kim, S.~Yang, J.~Kim, H.~Park, J.~Lee, and S.~Chang, ``Domain generalization with relaxed instance frequency-wise normalization for multi-device acoustic scene classification,'' in \emph{Interspeech}, 2022.

\bibitem{Schmid2022CPJKU22}
F.~Schmid, S.~Masoudian, K.~Koutini, and G.~Widmer, ``{CP-JKU} submission to {DCASE22}: Distilling knowledge for low-complexity convolutional neural networks from a patchout audio transformer,'' DCASE Challenge, Tech. Rep., 2022.

\bibitem{Morocutti23DIR}
T.~Morocutti, F.~Schmid, K.~Koutini, and G.~Widmer, ``Device-robust acoustic scene classification via impulse response augmentation,'' in \emph{EUSIPCO}, 2023.

\bibitem{Truchan25da}
H.~Truchan, T.~H. Ngo, and Z.~Ahmadi, ``Ascdomain: Domain invariant device-adversarial isotropic knowledge distillation convolutional neural architecture,'' in \emph{ICASSP}, 2025.

\bibitem{Koutini2020CPJKU20}
K.~Koutini, F.~Henkel, H.~Eghbal-zadeh, and G.~Widmer, ``{CP-JKU} submissions to {DCASE’20}: Low-complexity cross-device acoustic scene classification with {RF}-regularized {CNNs},'' DCASE Challenge, Tech. Rep., 2020.

\bibitem{Kim2021qti}
B.~Kim, S.~Yang, J.~Kim, and S.~Chang, ``{QTI} submission to {DCASE} 2021: Residual normalization for device-imbalanced acoustic scene classification with efficient design,'' DCASE Challenge, Tech. Rep., 2021.

\bibitem{Lee2022}
J.-H. Lee, J.-H. Choi, P.~M. Byun, and J.-H. Chang, ``Hyu submission for the {DCASE} 2022: Efficient fine-tuning method using device-aware data-random-drop for device-imbalanced acoustic scene classification,'' DCASE Challenge, Tech. Rep., 2022.

\bibitem{Koutini2021CPJKU21}
K.~Koutini, J.~Schlüter, and G.~Widmer, ``{CPJKU} submission to {DCASE21}: Cross-device audio scene classification with wide sparse frequency-damped {CNNs},'' DCASE Challenge, Tech. Rep., 2021.

\bibitem{Han2024rank1}
H.~Bing, H.~Wen, C.~Zhengyang, J.~Anbai, C.~Xie, F.~Pingyi, L.~Cheng, L.~Zhiqiang, L.~Jia, Z.~Wei-Qiang, and Q.~Yanmin, ``Data-efficient acoustic scene classification via ensemble teachers distillation and pruning,'' DCASE Challenge, Tech. Rep., 2024.

\bibitem{Yang2021lottery}
C.-H.~H. Yang, H.~Hu, S.~M. Siniscalchi, Q.~Wang, W.~Yuyang, X.~Xia, Y.~Zhao, Y.~Wu, Y.~Wang, J.~Du, and C.-H. Lee, ``A lottery ticket hypothesis framework for low-complexity device-robust neural acoustic scene classification,'' DCASE Challenge, Tech. Rep., 2021.

\bibitem{Tan2023a}
J.~Tan and Y.~Li, ``Low-complexity acoustic scene classification using blueprint separable convolution and knowledge distillation,'' DCASE Challenge, Tech. Rep., 2023.

\bibitem{Cai2023a}
Y.~Cai, M.~Lin, C.~Zhu, S.~Li, and X.~Shao, ``{DCASE2023} task1 submission: Device simulation and time-frequency separable convolution for acoustic scene classification,'' DCASE Challenge, Tech. Rep., 2023.

\bibitem{Schmid2023CPJKU23}
F.~Schmid, T.~Morocutti, S.~Masoudian, K.~Koutini, and G.~Widmer, ``{CP-JKU} submission to {DCASE23}: Efficient acoustic scene classification with cp-mobile,'' DCASE Challenge, Tech. Rep., 2023.

\bibitem{Shao2024rank2}
Y.-F. Shao, P.~Jiang, and W.~Li, ``Low-complexity acoustic scene classification with limited training data,'' DCASE Challenge, Tech. Rep., 2024.

\bibitem{Gemmeke17audioset}
J.~F. Gemmeke, D.~P.~W. Ellis, D.~Freedman, A.~Jansen, W.~Lawrence, R.~C. Moore, M.~Plakal, and M.~Ritter, ``Audio set: An ontology and human-labeled dataset for audio events,'' in \emph{ICASSP}, 2017.

\bibitem{Cai2024rank5}
Y.~Cai, M.~Lin, S.~Li, and X.~Shao, ``{DCASE2024} task1 submission: Data-efficient acoustic scene classification with self-supervised teachers,'' DCASE Challenge, Tech. Rep., 2024.

\bibitem{David2024rank2}
N.~David, R.~Aida, and S.~Patrick, ``Data-efficient acoustic scene classification with pre-trained {CP-Mobile},'' DCASE Challenge, Tech. Rep., 2024.

\bibitem{Werning2024rank8}
A.~Werning and R.~Haeb-Umbach, ``{Upb-Nt} submission to {DCASE24}: Dataset pruning for targeted knowledge distillation,'' DCASE Challenge, Tech. Rep., 2024.

\bibitem{Schmid2023cpm}
F.~Schmid, T.~Morocutti, S.~Masoudian, K.~Koutini, and G.~Widmer, ``Distilling the knowledge of transformers and {CNNs} with {CP}-mobile,'' in \emph{DCASE Workshop}, 2023.

\bibitem{Tan2025rank2}
E.-L. Tan, J.~W. Yeow, S.~Peksi, H.~Li, Z.~Yang, and W.-S. Gan, ``Sntl-ntu dcase25 submission: Acoustic scene classification using {CNN}-{GRU} model without knowledge distillation,'' DCASE2025 Challenge, Tech. Rep., May 2025.

\bibitem{Hu18SE}
J.~Hu, L.~Shen, and G.~Sun, ``Squeeze-and-excitation networks,'' in \emph{CVPR}, 2018.

\bibitem{Luo2025rank3}
Y.~Luo, H.~Liu, L.~Shi, and L.~Gan, ``Dynacp: Dynamic parallel selective convolution in cp-mobile under multi-teacher distillation for acoustic scene classification,'' DCASE2025 Challenge, Tech. Rep., 2025.

\bibitem{Chang2025rank5}
S.-G. Han, P.~M. Byun, and J.-H. Chang, ``Hyu submission for {DCASE} 2025 task 1: Low-complexity acoustic scene classification using reparameterizable {CNN} with channel-time-frequency attention,'' DCASE2025 Challenge, Tech. Rep., 2025.

\bibitem{Ramezanee2025rank7}
M.~M. Ramezanee, H.~Sharify, A.~M. Mehrani~Kia, and B.~Raoufi, ``Acoustic scene classification with knowledge distillation and device-specific fine-tuning for {DCASE} 2025,'' DCASE2025 Challenge, Tech. Rep., 2025.

\bibitem{Han25rank1_icassp}
B.~Han, W.~Huang, Z.~Chen, A.~Jiang, P.~Fan, C.~Lu, Z.~Lv, J.~Liu, W.-Q. Zhang, and Y.~Qian, ``Data-efficient low-complexity acoustic scene classification via distilling and progressive pruning,'' in \emph{ICASSP}, 2025.

\bibitem{zeng2025ctfa}
X.~Zeng and M.~Wang, ``Channel-time-frequency attention module for improved multi-channel speech enhancement,'' \emph{IEEE Access}, 2025.

\bibitem{Hinton15kd}
G.~E. Hinton, O.~Vinyals, and J.~Dean, ``Distilling the knowledge in a neural network,'' in \emph{NIPS Deep Learning Workshop}, 2014.

\bibitem{Li2025rank6}
H.~Li, Z.~Yang, M.~Wang, E.-L. Tan, J.~Yeow, S.~Peksi, and W.-S. Gan, ``Joint feature and output distillation for low-complexity acoustic scene classification,'' DCASE2025 Challenge, Tech. Rep., 2025.

\bibitem{Jeong2025rank8}
S.~Jeong and S.~Kim, ``Adaptive knowledge distillation using a device-aware teacher for low-complexity acoustic scene classification,'' DCASE2025 Challenge, Tech. Rep., 2025.

\bibitem{Chen2025rank9}
X.~Chen and W.~Xie, ``{McCi} submission to {DCASE} 2025: Training low-complexity acoustic scene classification system with knowledge distillation and curriculum,'' DCASE2025 Challenge, Tech. Rep., 2025.

\bibitem{koutini22passt}
K.~Koutini, J.~Schl{\"{u}}ter, H.~Eghbal{-}zadeh, and G.~Widmer, ``Efficient training of audio transformers with patchout,'' in \emph{Interspeech}, 2022.

\bibitem{chen23beats}
S.~Chen, Y.~Wu, C.~Wang, S.~Liu, D.~Tompkins, Z.~Chen, W.~Che, X.~Yu, and F.~Wei, ``{BEATs}: Audio pre-training with acoustic tokenizers,'' in \emph{ICML}, 2023.

\bibitem{Schmid23EfficientAT}
F.~Schmid, K.~Koutini, and G.~Widmer, ``Efficient large-scale audio tagging via transformer-to-cnn knowledge distillation,'' in \emph{ICASSP}, 2023.

\bibitem{Schmid24DyMN}
------, ``Dynamic convolutional neural networks as efficient pre-trained audio models,'' \emph{{IEEE} {ACM} Trans. Audio Speech Lang. Process.}, 2024.

\bibitem{Karasin2025rank1}
D.~Karasin, I.-C. Olariu, M.~Schöpf, and A.~Szymańska, ``Domain-specific external data pre-training and device-aware distillation for data-efficient acoustic scene classification,'' DCASE2025 Challenge, Tech. Rep., May 2025.

\bibitem{CochlScene}
I.-Y. Jeong and J.~Park, ``Cochlscene: Acquisition of acoustic scene data using crowdsourcing,'' in \emph{APSIPA ASC}, 2022.

\bibitem{koutini21cpr}
K.~Koutini, H.~Eghbal{-}zadeh, and G.~Widmer, ``Receptive field regularization techniques for audio classification and tagging with deep convolutional neural networks,'' \emph{{IEEE} {ACM} Trans. Audio Speech Lang. Process.}, 2021.

\bibitem{Han2025rank12}
S.~Han, D.~H. Lee, M.~S. Jo, E.~S. Ha, M.~J. Chae, and G.~W. Lee, ``Confidence-aware ensemble knowledge distillation for low-complexity acoustic scene classification,'' DCASE2025 Challenge, Tech. Rep., 2025.

\end{thebibliography}

%
%
%
%
%
%
%
%
%

\end{sloppy}

\end{document}